\documentclass[conference]{IEEEtran}
\IEEEoverridecommandlockouts

\usepackage{tabularx}
\usepackage{array}
\usepackage{makecell}
\usepackage[utf8]{inputenc}
\usepackage{amsmath}
\usepackage{amssymb}
\usepackage{datetime}
\usepackage{fancyhdr}
\usepackage[margin=2cm]{geometry}
\usepackage{latexsym}
\usepackage{nth}
\usepackage{graphicx}
\usepackage{url}
\usepackage{svg}
\usepackage{wrapfig}
\usepackage{float}
\usepackage{subcaption}
\usepackage{multirow}
\usepackage{stfloats}
\usepackage{cite}

\usepackage{booktabs}
\usepackage[hidelinks]{hyperref}

\usepackage{titlesec}

\begin{document}

\title{Transformer Redesign for Late Fusion of Audio-Text Features on Ultra-Low-Power Edge Hardware\\
\thanks{${\dagger}$ These authors contributed equally to this work.}
}

\author{
    \IEEEauthorblockN{
        Stavros Mitsis$^{\dagger}$, 
        Ermos Hadjikyriakos$^{\dagger}$, 
        Humaid Ibrahim$^{\dagger}$,\\
        Savvas Neofytou$^{\dagger}$, 
        Shashwat Raman$^{\dagger}$, 
        James Myles$^{\dagger}$, 
        Eiman Kanjo$^{\dagger\ddagger}$
    }
    \IEEEauthorblockA{
        $^{\dagger}$Department of Computing, Imperial College London, United Kingdom\\
        \{stavros.mitsis24, ermos.hadjikyriakos24, humaid.ibrahim24,\\
        savvas.neofytou24, shashwat.raman24, james.myles24, e.kanjo\}@imperial.ac.uk\\
    }
    \IEEEauthorblockA{
        $^{\ddagger}$Nottingham Trent University, Nottingham, United Kingdom\\
        eiman.kanjo@ntu.ac.uk
    }
}

\maketitle

\begin{abstract}
Deploying emotion recognition systems in real-world environments where devices must be small, low-power, and private remains a significant challenge. This is especially relevant for applications such as tension monitoring, conflict de-escalation, and responsive wearables, where cloud-based solutions are impractical. 
Multimodal emotion recognition has advanced through deep learning, but most systems remain unsuitable for deployment on ultra-constrained edge devices. Prior work typically relies on powerful hardware, lacks real-time performance, or uses unimodal input. This paper addresses that gap by presenting a hardware-aware emotion recognition system that combines acoustic and linguistic features using a late-fusion architecture optimised for Edge TPU. The design integrates a quantised transformer-based acoustic model with frozen keyword embeddings from a DSResNet-SE network, enabling real-time inference within a 1.8~MB memory budget and 21--23~ms latency. The pipeline ensures spectrogram alignment between training and deployment using MicroFrontend and MLTK. Evaluation on re-recorded, segmented IEMOCAP samples captured through the Coral Dev Board Micro microphone shows a 6.3\% macro F1 improvement over unimodal baselines. This work demonstrates that accurate, real-time multimodal emotion inference is achievable on microcontroller-class edge platforms through task-specific fusion and hardware-guided model design.
\end{abstract}


\section{Introduction}
Recent advancements in speech and language processing have notably enhanced our capability to analyze human communication; however, applying these technologies to detect critical social issues, such as conflict and abusive behavior, remains challenging due to significant computational demands. High-performing machine learning models typically necessitate substantial processing resources, restricting their deployment primarily to cloud-based infrastructures or expensive hardware. Consequently, economically disadvantaged communities, who could most benefit from timely intervention and protective technologies, often lack access to these critical tools \cite{deboer2024using}.

Motivated by leveraging artificial intelligence (AI) for social good, particularly in domestic violence detection through conversational analysis, this research prioritises the development of lightweight AI architectures optimized for resource-constrained microcontroller units (MCUs). Deploying efficient models directly on affordable edge devices democratizes AI, making advanced analytical tools broadly accessible regardless of economic status, thereby reducing dependency on costly computational infrastructure \cite{malik2023emotions}.

Privacy preservation represents another crucial research driver. Conventional high-performance AI solutions rely predominantly on centralized cloud computing, necessitating the external transmission and processing of personal data, thereby exacerbating privacy and security concerns. In contrast, deploying models locally on MCUs ensures users maintain full control over sensitive data, significantly enhancing privacy and security \cite{ali2021privacy, tsouvalas2022privacy}. \\
\indent Initially, the project aimed to directly detect domestic violence within conversational speech. Due to the scarcity of sufficiently large and specifically annotated violent interaction datasets, the research scope pivoted strategically toward general emotion classification. This approach leveraged established emotional speech datasets while adhering to core technical objectives: lightweight architecture, modular design, and real-time inference on low-power devices. Critically, the modularity of the proposed system combining keyword detection with comprehensive spectrogram acoustic analysis facilitates future adaptation to violence detection tasks pending dataset availability. \\
\indent Integrating targeted keyword detection and detailed spectrogram analysis, the developed multi-level fusion model effectively captures both semantic content and paralinguistic nuances. This dual-modality framework markedly enhances sensitivity to subtle emotional expressions and potential linguistic indicators of abuse. Furthermore, the final system implementation employs model quantisation and architectural compression, enabling efficient real-time inference of emotions (negative emotions) on low-cost, resource-limited MCUs.

%%%%%%%%%%%%%%%%%%%%%%%%%%%
%%%%%%%%%%%%%%%%%%%%%%%%%%%

\section{Background and Related Work}

\subsection{Keyword Detection in Speech Processing}

Keyword spotting (KWS) is the task of detecting predefined keywords or phrases from continuous speech, often in real-time and under resource constraints. It is a core component of many voice-activated systems, such as wake-word detection in virtual assistants like Apple's “Hey Siri” or Google's “OK Google” \cite{apple2017heysiri, ok_google}. KWS has long been recognized as a critical task in speech processing \cite{rose1990_kwsold, wilpon1990_kwsold, rohlicek1989_kwsold}, and has evolved significantly from early statistical techniques to modern neural network-based approaches. Traditional systems often relied on Hidden Markov Models (HMMs) and Gaussian Mixture Models (GMMs), but have since transitioned to more powerful and efficient end-to-end deep learning models that better capture spectral and temporal features of speech. Early deep learning-based approaches used fully connected Deep Neural Networks (DNNs), which were later surpassed by convolutional and recurrent architectures for improved accuracy and efficiency on streaming audio \cite{zhang2018, choi2019}.

% Keyword spotting (KWS) is a well-established task in speech processing, and its techniques have evolved significantly from early statistical methods to modern deep learning models. KWS has transitioned from traditional Hidden Markov Model pipelines to end-to-end neural models that capture both spectral and temporal patterns. Early deep learning efforts used fully connected deep neural networks (DNNs), but these have largely been replaced by convolutional and temporal architectures that deliver better accuracy and efficiency on streaming audio \cite{zhang2018, choi2019}.

Depthwise separable convolutional neural networks (DS-CNNs) like MobileNet \cite{andrew2017}, factorise standard convolutions into lightweight depthwise and pointwise operations, reducing parameter count and compute. Zhang et al. demonstrated that such DS‐CNNs outperform similarly sized DNNs by roughly 10\% in accuracy on Google’s Speech Commands benchmark \cite{zhang2018}. Temporal convolutional networks (TCNs) and compact ResNet‐based variants similarly exploit convolutional bottlenecks and residual connections for low-latency inference on-device \cite{li2020, choi2019}.

Channel-wise attention via Squeeze-and-Excitation (SE) blocks further refines feature representations with minimal overhead. SE‐enhanced CNNs have repeatedly shown small but consistent gains in KWS accuracy: for example, a depthwise separable ResNet augmented with SE modules outperformed baseline CNNs at comparable model sizes \cite{xu2020, hu2019}. More recently, broadcasted residual networks (BC-ResNet) architectures replace most 2D filters with 1D time-domain convolutions and “broadcast” outputs across frequency bands, achieving state-of-the-art accuracy (over 98\% on 35 commands) with fewer operations \cite{kim2023}.

% Beyond pure convolution, recurrent and attention-based models have been explored for richer temporal modelling. Bi-directional LSTMs with attention mechanisms capture long-range dependencies in continuous speech and can outperform CNNs when latency constraints permit \cite{rai2023}. Fully self-attentional architectures, such as the Keyword Transformer (KWT), replace all convolutions with multi-head attention and have set new accuracy records (e.g.\ ~98.6\% on 12-command tasks) without pre-training, albeit with higher computational demands \cite{berg2021}. These developments highlight the trade-off between compactness for edge deployment and the ability to model global context for maximal accuracy.
\subsection{Sentiment Analysis Using Audio}
Emotion recognition from speech and text has been widely studied using both unimodal and multimodal approaches, with the IEMOCAP dataset emerging as a popular benchmark.

In unimodal speech emotion recognition (SER), early systems relied heavily on handcrafted features such as pitch, energy, and Mel-Frequency Cepstral Coefficients (MFCCs). However, recent methods have shifted toward deep learning architectures. CNNs and long short-term memory (LSTM) networks have proven effective in capturing both spectral and temporal features from speech spectrograms \cite{Yang2018, Satt2017}. More recently, self-supervised learning models like wav2vec 2.0 have achieved remarkable performance by leveraging large pretraining corpora and fine-tuning on smaller emotion-labeled datasets \cite{Morais2021, Makiuchi2021}. Capsule networks (CapsNets) have also gained traction by preserving spatial relationships within spectrograms, enhancing feature representation \cite{Wu2019}.

Multimodal systems, which integrate both acoustic and linguistic modalities, have shown further improvements in emotion classification. Late fusion approaches combining scores from BERT-based text classifiers and CNN or ResNet-based speech models allow leveraging complementary strengths of each modality \cite{Padi2022}. Similarly, feature-level fusion using embeddings from wav2vec 2.0 and Transformers has yielded strong results \cite{Makiuchi2021}. Multi-task learning (MTL) has also proven effective, where models simultaneously perform speech-to-text and emotion classification tasks, leading to better generalisation and shared feature representations \cite{Cai2021}.

% Together, these advancements highlight the field's trajectory toward more robust, data-efficient, and generalizable emotion recognition systems. The consistent use of the IEMOCAP dataset across studies further reinforces its utility and importance in benchmarking progress in SER research.

% \subsection{Edge Computing and Quantization}
% \begin{itemize}
%     \item Challenges of deploying deep neural networks on edge devices
%     \item Overview of quantization techniques for model optimization
% \end{itemize}
% \input{Edge_Computing_and_Quantization}

%%%%%%%%%%%%%%%%%%%%%%%%%%%
%%%%%%%%%%%%%%%%%%%%%%%%%%%

\section{Datasets}

% The three primary datasets employed are the IEMOCAP corpus for emotional speech analysis, a curated dataset for KWS using LibriSpeech, and the MUSAN noise corpus for augmentation, which support a comprehensive evaluation of emotional, lexical, and acoustic variations.

\subsection{IEMOCAP Dataset}
% \begin{itemize}
%     \item Overview of the dataset (emotional speech samples)
%     \item Relevance for training and evaluating sentiment analysis models
%     \item Comparison with LibriSpeech in terms of diversity and acoustic properties
% \end{itemize}
%
The IEMOCAP dataset is a multimodal corpus developed by the University of Southern California~\cite{busso2008iemocap}. It consists of 12 hours of richly annotated audiovisual recordings, capturing speech audio, video streams, facial motion capture data, and textual transcriptions. This dataset is structured into five distinct recording sessions, each involving a pair of professional actors (totalling 10 unique speakers—five males and five females). These sessions consist of both scripted dialogues and improvised conversations, intentionally designed to elicit and cover a diverse range of emotional expressions.

Each dialogue segment within IEMOCAP is annotated by expert evaluators, ensuring label reliability. Annotations include categorical labels such as anger, happiness, sadness, frustration, excitement, and neutral. Audio recordings within IEMOCAP are provided in stereo format, with each channel capturing one speaker separately, thus facilitating targeted speaker-specific emotion analysis. 

Various state-of-the-art emotion recognition approaches have been evaluated using IEMOCAP, achieving noteworthy performance metrics~\cite{wagner2022dawn}~\cite{xu2021multiscale}. For example, certain methodologies have reported weighted accuracies surpassing 75 percent for categorical emotion classification tasks, underscoring the dataset's efficacy and significance in driving forward research in emotional computing and machine learning applications.

% \subsubsection{Recording of Test Dataset}
For training and validation purposes, under conditions that reflect the actual audio quality of the microcontroller, one session from the original dataset was re-recorded using the microcontroller's microphone. This approach ensured the development of a test and validation dataset with identical microphone characteristics to those of the target hardware.

To guarantee consistency and reproducibility, the following experimental setup was employed. A custom stand was designed and 3D printed to securely hold the microcontroller in a fixed position throughout the recording process. Audio playback was delivered via a 2020 MacBook Pro, with the microcontroller positioned equidistantly from the laptop's stereo speakers, at an approximate distance of 15 cm from each speaker.

% A render of this stand is shown in Figure~\ref{fig:stand}. 
% \begin{figure}
%     \centering
%     \includegraphics[width=0.5\linewidth]{microcontroller/figures/microcontroller_stand.jpg}
%     \caption{Render of Custom Microcontroller Stand}
%     \label{fig:stand}
% \end{figure}

The microcontroller was programmed with custom C\texttt{++} firmware to enable real-time audio streaming to a Linux-based laboratory workstation. A Python script on the workstation captured the audio stream and saved it in \texttt{.wav} format. Following the recording, the audio was compressed and converted to \texttt{.mp3} format for storage and further processing. The recording session lasted a total of 2 hours, 17 minutes, and 38 seconds at a sampling rate of 16 kHz. The resulting audio file was subsequently segmented into individual clips that correspond precisely to those in the original dataset.

\subsection{LibriSpeech Dataset}

A dataset of spoken keywords and non-keyword audio was curated from open-source corpora. The primary source for the keywords was the LibriSpeech corpus \cite{panayotov2015}, which contains roughly 1000 hours of read English speech from audiobooks sampled at 16 kHz. From LibriSpeech, utterances of our target keywords that are common in IEMOCAP dataset were extracted. The start and end timings of the words are extracted using the Montreal Forced Aligner \cite{mcauliffe17_montreal}, similarly to Bittar et al. \cite{bittar2024}. Using the provided alignments, the audio was segmented so each clip contains a single keyword.

\subsection{MUSAN Dataset}
To increase noise robustness, the MUSAN corpus~\cite{snyder2015} was incorporated into the datasets. The noise examples consist of recordings such as construction noises, crowd chatter, music, and more. These partitions were mixed with LibriSpeech and IEMOCAP utterances to emulate real-world acoustic conditions, a strategy shown to reduce over-fitting and false-alarm rates in noisy environments~\cite{yang23}.

\section{Microcontroller}
The Coral Micro DevBoard was selected due to its compact footprint (65$\times$30), minimal RAM capacity (64 MB), and dual-core ARM Cortex CPU combined with a dedicated Tensor Processing Unit (TPU) for accelerated on-device machine learning (ML) inference. As such the Coral Micro DevBoard is particularly suitable for deployment in resource-constrained environments.\cite{coral2025microdatasheet}

\subsection{Software Ecosystem: Coral Micro Repository}
Google's Coral Micro repository \cite{coral2025github}, developed in C++, provides a comprehensive software stack including the FreeRTOS operating system \cite{freertos2025fundamentals}, essential libraries, device drivers, memory management utilities, and a custom-optimized TensorFlow Lite Micro (TFLM) framework for resource-constrained machine learning inference.

Key libraries within the repository include: \texttt{Kiss FFT}, a lightweight Fast Fourier Transform (FFT) library for spectral analysis; \texttt{TF-Lite MicroFrontend}, an advanced audio feature extraction pipeline incorporating Kaiser windowing, FFT, log-energy compression, Per-Channel Energy Normalization (PCAN) auto-gain, and noise reduction; \texttt{AudioDriver}, a dedicated C++ interface for the onboard Pulse-Density Modulation (PDM) microphone; and \texttt{edgetpu\_compiler}, which converts TensorFlow Lite Micro models into a TPU-compatible format to enable efficient on-device inference acceleration.

%%%%%%%%%%%%%%%%%%%%%%%%%%%%%%%%%%%%%%%%%%%%%%%%%%%%%%%%%%%
% IEE REPORT CONTENT 
%%%%%%%%%%%%%%%%%%%%%%%%%%%%%%%%%%%%%%%%%%%%%%%%%%%%%%%%%%%

\subsection{Mel-Spectrograms Generation}
\label{sec:spectrogram}
 Time–frequency representations such as Mel-spectrograms are widely used in audio processing. A Mel-spectrogram is created by applying a short-time Fourier transform to the waveform, mapping the resulting linear-frequency bins onto the Mel scale, and logarithmically compressing the band energies. These 2D representations serve as the primary input to the keyword and emotion classification models in this study and are also standard in other domains, including environmental sound tagging, keyword spotting, and speech-to-text systems such as OpenAI's Whisper \cite{radford2022}.

\subsection{Spectrogram Alignment: Training - Inference}

To ensure consistency between spectrogram creation libraries during training (Python/ \texttt{Librosa}) and on-device inference (Coral Micro’s \texttt{MicroFrontend}), an alignment process was required due to functional and argument discrepancies between the two libraries. While \texttt{Librosa} offers extensive configurability and feature extraction capabilities, \texttt{MicroFrontend} on Coral Micro provides a fixed, resource-efficient pipeline, including predefined windowing, padding modes, and log-energy compression using lookup tables and bit shifts. Resolving these differences was critical to maintain identical feature representations during both training and inference.

Initial attempts to modify \texttt{MicroFrontend} or port \texttt{LibrosaCpp} to the microcontroller were unsuccessful due to cascading runtime failures, memory overflows, and excessive computational delays that rendered real-time inference infeasible. To overcome these limitations, the Silicon Labs Machine Learning Toolkit (\texttt{MLTK}), a Python wrapper providing full access to \texttt{MicroFrontend}'s internal configuration, was adopted. \texttt{MLTK} enabled seamless alignment of spectrogram generation across both environments.

\subsection{Final Configuration and Empirical Validation}
\begin{figure}[htbp]
  \centering
  \begin{subfigure}[b]{0.4\textwidth}
    \centering
    \includegraphics[width=\linewidth]{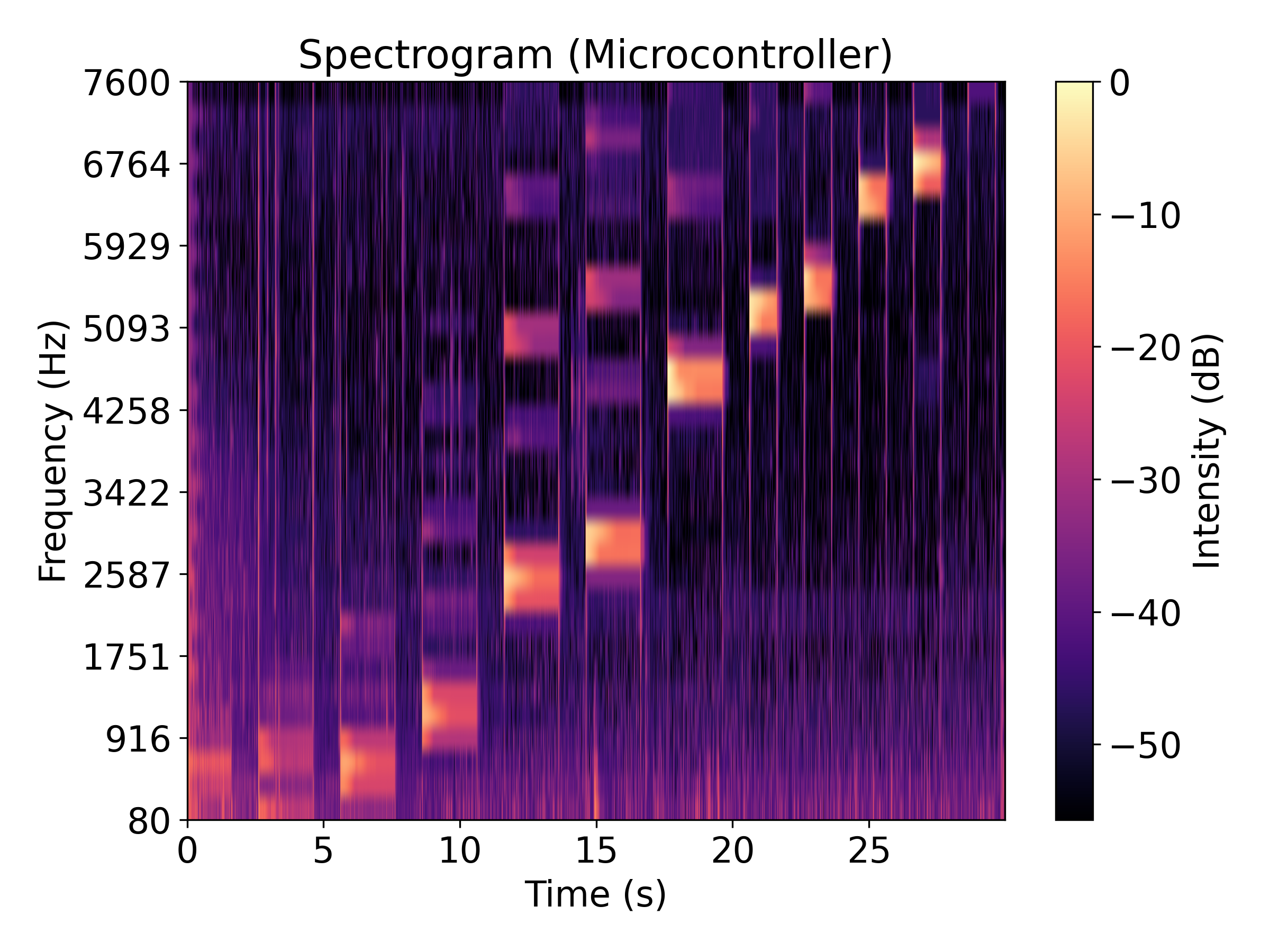}
    \caption{Spectrogram computed on Coral Micro.}
    \label{fig:spectrogram_coral}
  \end{subfigure}
  \hfill
  \begin{subfigure}[b]{0.4\textwidth}
    \centering
    \includegraphics[width=\linewidth]{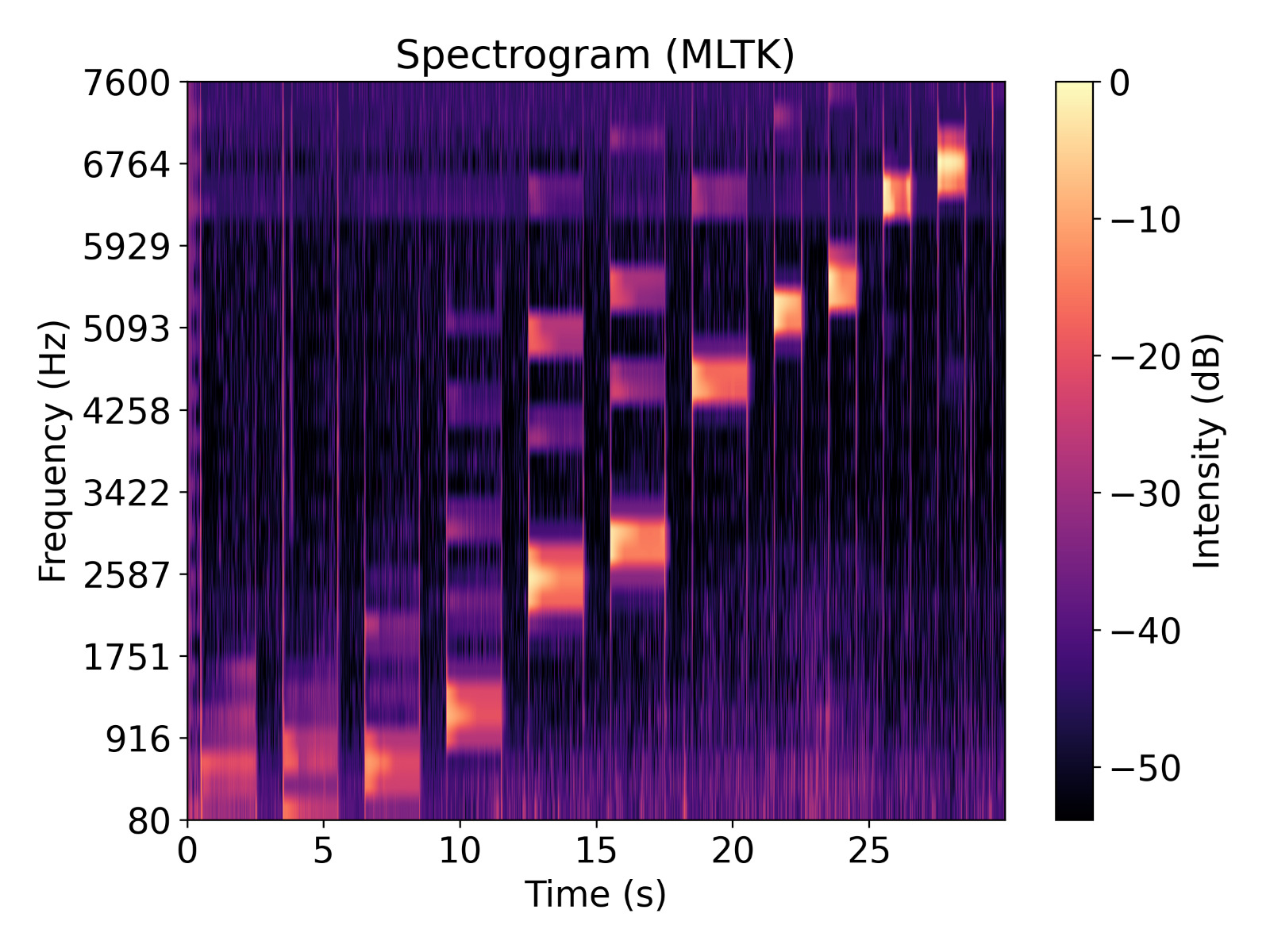}
    \caption{Spectrogram computed by \texttt{MLTK} on PC.}
    \label{fig:spectrogram_mltk}
  \end{subfigure}
  \caption{Comparison of mel-spectrograms: (a) generated on Coral Micro, and (b) re-recorded as \texttt{.wav} and processed using \texttt{MLTK}. The horizontal axis denotes time (s), and the vertical axis denotes frequency (Hz); color indicates intensity in dB.}
  \label{fig:spectrogram_comparison}
\end{figure}

End-to-end consistency was validated by generating a 30-second, 16 kHz \texttt{.wav} file containing sequential tonal bursts. The audio was replayed twice using an iPhone 16 Pro Max positioned 10cm from the Coral Micro's microphone. In the first pass, the device saved the recorded audio for offline spectrogram generation via \texttt{MLTK}; in the second pass, the audio was processed in real time by \texttt{MicroFrontend}, producing a spectrogram output. Accounting for minor ambient noise variations, the comparison confirms alignment between training and inference pipelines (Figure~\ref{fig:spectrogram_comparison}).

The final settings for both the \texttt{MLTK} and \texttt{MicroFrontend}, and for all experiments performed in the study were: sample rate 16 kHz; 25 ms window with 10 ms hop; 32 Mel channels spanning 80–7,600 Hz; noise reduction enabled (smoothing bits = 10, even = 0.025, odd = 0.06, min‐signal = 0.05); log‐scale shift = 6. As an additional adaptation, the audio initially loaded into the training pipeline is kept in 16-bit integer (\texttt{INT16}) format, the exact raw PCM output produced by the Coral Micro’s PDM microphone. This is to ensure that the downstream feature extractor sees identical amplitude dynamics and quantization behaviour both on and outside the Coral Micro environment.

\section{Multi-modal Emotion Classification Architecture}
\begin{figure}[H]
    \centering
    \includegraphics[width=\linewidth]{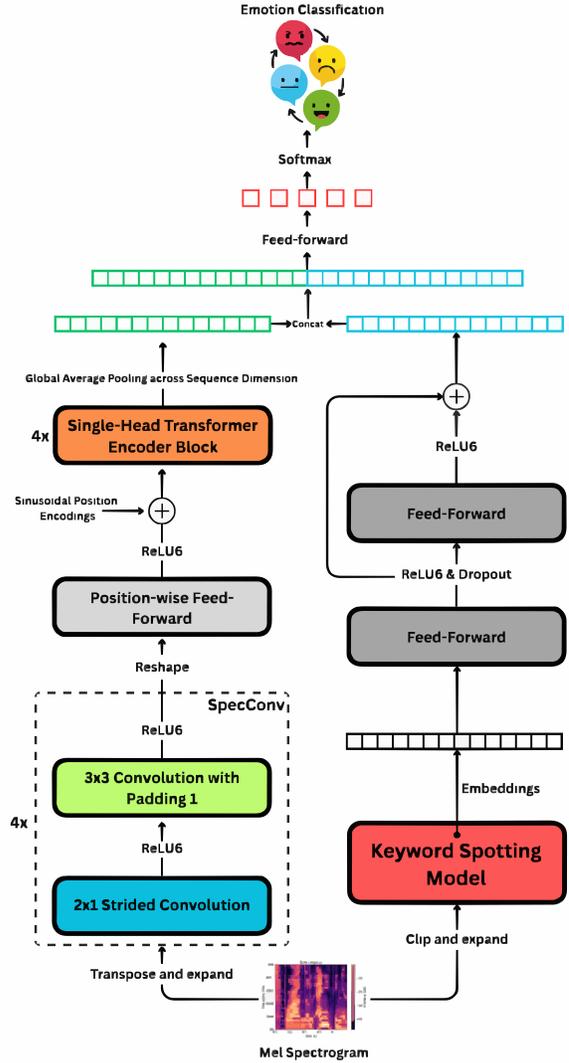}
    \caption{System Architecture}
    \label{fig:sys_arch}
\end{figure}

% late fusion multi-modal architectur
The emotion classification architecture (Figure \ref{fig:sys_arch}) is composed of two main components: a textual feature extractor, which contains a lightweight keyword spotting (KWS) model, and a ViT-based \cite{dosovitskiy2020image} audio feature extractor model. Rather than being used directly for classification, these models act as rich feature extractors, each producing a modality-specific embedding. The two embeddings are processed by the classification head to produce a final emotion classification. This late-fusion design enables the system to perform robust multi-modal inference while maintaining low latency and power consumption. \texttt{ReLU6} activation functions are used throughout the model in order to maximise post-quantization performance \cite{Krizhevsky2010}, and dropout is applied to aid generalisation.
% \begin{figure}[H]
%     \centering
%     \includesvg[width=\linewidth]{sentiment-analysis-model/figures/System_Architecture.svg}
%     \caption{System Architecture}
%     \label{fig:sys_arch}
% \end{figure}

\subsection{ViT-based Audio Feature Extractor}
In order to capture both long- and short-range dependencies between frames in the spectrogram, a light-weight ViT encoder ~\cite{dosovitskiy2020image} is deployed. Specifically, after transposing the spectrogram, a CNN reduces the temporal dimension, significantly reducing the number of parameters required by the attention modules in the transformer blocks.

% \begin{wrapfigure}[37]{r}{0.5\textwidth}
%     \centering
%     \includesvg[width=\linewidth]{sentiment-analysis-model/figures/System Architecture.svg}
%     \caption{System Architecture}
%     \label{fig:sys_arch}
% \end{wrapfigure}

\indent The CNN consists of $4$ blocks, denoted \texttt{SpecConv}, each containing a strided $2 \times 1$ convolution to condense information from nearby time-steps, and a $3 \times 3$ convolution with padding $1$, which preserves spatial dimensions. A fully-connected layer is applied row-wise to the output of the CNN, projecting tokens to dimension $d_{model}$. \newline
\indent Sinusoidal positional encodings are injected to the outputs of the CNN, before tokens are propagated through a sequence of $4$ transformer-based encoder blocks \cite{transformer}. These blocks contain a self-attention mechanism, a position-wise feed-forward neural network, layer-normalisation and residual connections. Finally, global average pooling is applied across the sequence dimension to produce an encoding of length $d_{model}$, which is received by the classification head.
%These embeddings are then combined with the passed into a secondary emotion classification model, which is tasked with detecting emotional states. The emotion classifier 
\subsection{Textual Feature Extractor}
The primary component of the textual feature extractor is a lightweight KWS model, described in Section \ref{sec:keyword_architecture}, which is trained separately and kept frozen whilst training the emotion classification model. Spectrograms are clipped to match the input size of the KWS model, which produces an embedding that captures linguistic features within the spectrogram. Specifically, activations from the penultimate layer of the KWS model are extracted, resulting in a $256$-dimensional embedding. A block consisting of two fully-connected layers, interleaved with dropout, \texttt{ReLU6} activation functions and a residual connection, is applied to extract information from the embedding and convert its size to $d_{model}$, before it is passed to the classification head.
\begin{table}[ht]
    \centering
    \scriptsize
    \begin{tabular}{llcc}
        \toprule
        \textbf{Module} & \textbf{Input Size} & \textbf{Operation} & \textbf{\# Params.} \\
        \midrule
        \multirow{4}{*}{\shortstack{Keyword\\Encoder}} 
            & 32×498        & Clip and expand                            & 0       \\
            & 32×490×1      & Extract Keyword Embedding                   & 194,853 \\
            & 256           & Lin. $256 \rightarrow 128$, ReLU6, Dropout & 32,896 \\
            & 128           & Lin. $128 \rightarrow 128$, ReLU6         & 16,512 \\
        \midrule
        \multirow{8}{*}{\shortstack{Spectrogram\\Encoder}}
            & 32×498        & Transpose and expand                        & 0       \\
            & 498×32×1      & SpecConv (16 filters)                       & 2,368   \\
            & 32×249×16     & SpecConv (32 filters)                       & 10,304  \\
            & 32×125×32     & SpecConv (64 filters)                       & 41,088  \\
            & 32×63×64      & SpecConv (1 filter)                         & 139     \\
            & 32×32         & Lin. $32 \rightarrow 128$                & 4,224   \\
            & 32×128        & Positional Encodings                        & 0       \\
            & 32×128        & Transformer Block ×4                        & 99,584×4 \\
            & 32×128        & AvgPool2D $32 \times 1$                     & 0       \\
        \midrule
        \multirow{3}{*}{\shortstack{Classification\\Head}}
            & 128 + 128     & Concatenate                                 & 0       \\
            & 256           & Lin. $256 \rightarrow 128$, ReLU6, Dropout & 32,896 \\
            & 256           & Lin. $128 \rightarrow 5$, Softmax         & 645     \\
        \bottomrule
    \end{tabular}
    \caption{Emotion Classification Architecture. 'Lin.' represents a linear layer operation}
    \label{tab:emotion_model_architecture}
\end{table}
\subsection{Classification head}
The classification head receives the acoustic and audio embeddings corresponding to the outputs of the audio and textual feature extractors and concatenates them into a single vector. Finally, a fully-connected layer and \texttt{Softmax} is applied, converting logits to a probability distribution over the five emotion categories.
\subsection{Keyword Spotting Model Architecture}
% \vspace{-0.4cm}
\label{sec:keyword_architecture}

% \begin{figure}[htbp]
%     \centering
%     \includesvg[width=0.5\textwidth]{keyword/figures/Keyword Model Architecture.svg}
%     \caption{Keyword model architecture, with a magnified residual block}
%     \label{fig:keyword_and_res_block}
% \end{figure}

% % The keyword spotting (KWS) model developed in this study is a compact convolutional neural network tailored for TinyML-class devices. Its architecture draws inspiration from MobileNet-like efficient CNNs and modern squeeze-and-excitation networks. As this model will be used as a feature extractor for a larger model, making it lightweight is of the utmost importance. For this reason, attention mechanisms are not suitable for this task, despite their accuracy in the KWS literature. Additionally, the attention mechanism is not natively supported on the microcontroller's accelerator chip, the Edge TPU. Newer architectures like BC-ResBlock \cite{kim2023}, rely on layers not present in the Edge TPU supported operations list \cite{coral2025edgetpu, coral_paper}, like SubSpectralNorm, making them difficult to implement on the microcontroller. 

% add a part on naming the model DSResNet-SE

% \begin{wrapfigure}[25]{r}{0.35\textwidth}
%     \centering
%     \includesvg[width=\linewidth]{keyword/figures/KWSModel2.svg}
%     \caption{DSResNet-SE model architecture, with a magnified residual block}
%     \label{fig:keyword_and_res_block}
% \end{wrapfigure}

\begin{figure*}[t]
    \centering
    \includegraphics[width=0.9\linewidth]{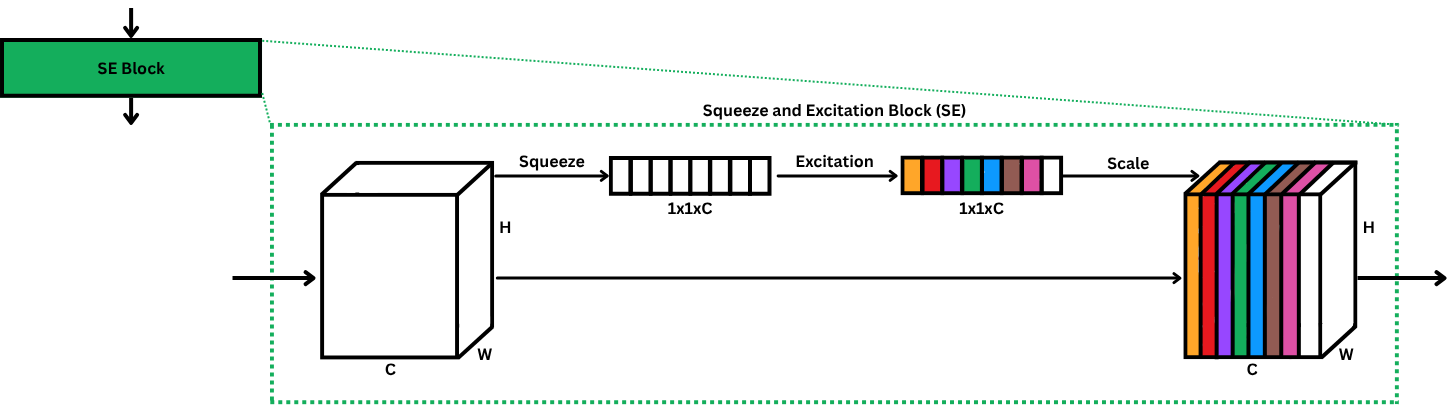}
    % \includesvg[width=0.8\linewidth]{keyword/figures/SE Architecture.svg}
    \caption{Squeeze and Excitation (SE) block architecture}
    \label{fig:enter-label}
\end{figure*}

\begin{figure}[H]
    \centering
    \includegraphics[width=0.9\linewidth]{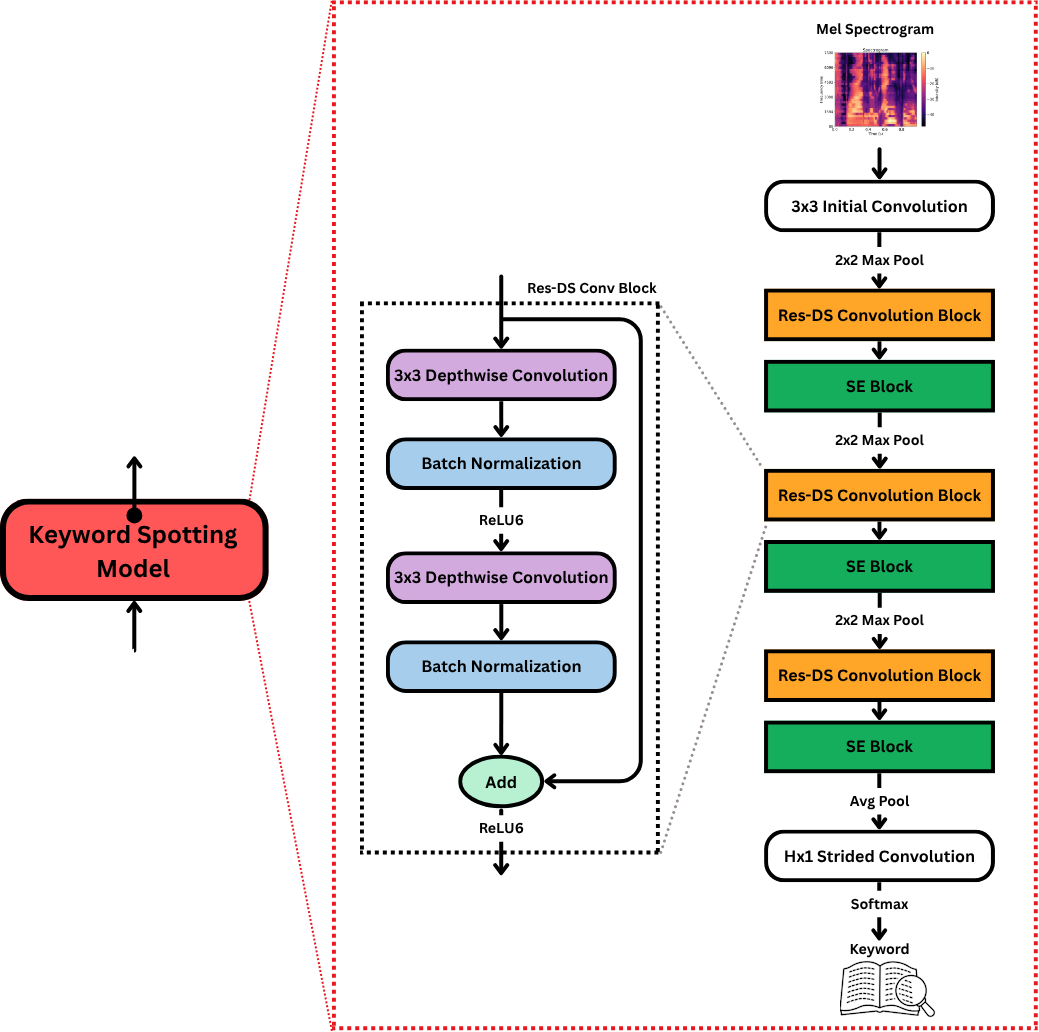}
    % \includesvg[width=\linewidth]{keyword/figures/KWSModel2.svg}
    \caption{DSResNet-SE model architecture, with a magnified residual block}
    \label{fig:keyword_and_res_block}
\end{figure}

The keyword spotting model, named DSResNet-SE, is a compact CNN designed based on ResNet and Squeeze-and-Excitation (SE) architectures, heavily inspired by Xu et al. \cite{xu2020}. Building upon the depthwise separable convolutional ResNet blocks, it combines them with SE blocks to act as a lightweight attention mechanism. Newer blocks such as broadcasted ResNets (BC-ResBlock) \cite{kim2023} were omitted as they depend on operations (e.g. SubSpectralNorm) absent from the Edge TPU’s supported list \cite{coral2025edgetpu, coral_paper}.

With an initial 32-filter convolution, the model uses repeated depthwise-separable convolutional blocks, each comprising a depthwise convolution, pointwise 1 × 1 convolution, batch normalisation, and a \texttt{ReLU6} activation for quantization efficiency \cite{Krizhevsky2010}. The pointwise 1x1 convolution greatly reduces the parameter count compared with a standard 2D convolution without harming audio performance \cite{zhang2018}. Chen et al. found that residual design stabilises training of deeper networks and boosts KWS accuracy by improving gradient flow and encouraging feature reuse \cite{chen2019}.

After each residual block, SE blocks were applied \cite{hu2019}, using global average pooling to squeeze each channel to a scalar and a small gating network to reweigh them, effectively serving as a lightweight attention mechanism. \texttt{Softmax} was applied to generate class-probability vectors for each sub-window. The feature embeddings for the emotion classification model were generated from the pooled output of the last SE block. These architectural choices keep the network lightweight under resource constrained environments.

In the literature, KWS models typically take in a 1-second input spectrogram \cite{xu2020, kim2023, yang23}. However, as the acoustic encoder model uses 5-second inputs, and due to the Edge TPU's inability to process batched inputs (mentioned in section \ref{edgetpu_limitations}), the network instead ingests a 5-second input. Five random 1-second audio buffers have been stitched together and converted to a single Mel-spectrogram. Because this 5-second window is equivalent to five consecutive 1-second clips, the model is designed to output five parallel predictions (one per second) in a single forward pass.
% \subsection{Emotion Classification Model}
% \begin{itemize}
%     \item Architecture of the secondary CNN/Transformer model
%     \item Process for extracting keyword embeddings from the frozen model
%     \item How the combined features are processed to predict sentiment
%     \item Discussion of the design choices made for multi-modal fusion
%     \item Overview of fusion strategies: concatenation, attention mechanisms, or transformer-based modules
% \end{itemize}
%%%%%%%%%%%%%%%%%%%%%%%%%%%
%%%%%%%%%%%%%%%%%%%%%%%%%%%

\section{Experimental Methodology}

\subsection{KWS Model}
\subsubsection{Dataset Preprocessing} \label{kw_preprocessing}

A list of target words were collected from the IEMOCAP transcripts. Stop words and words outside the English vocabulary were removed (e.g. country abbreviations). The top 100 words per emotion were extracted and filtered leading to a list of 58 target keywords. Using the alignments generated by Montreal Forced Aligner, one-second clips were extracted. Keywords with fewer than 2,000 instances were dropped, and those exceeding 20,000 were downsampled, resulting in a final list of 49 keywords.

% Due to microcontroller limitations, five 1-second audio clips had to be combined into one. The five clips were taken randomly, and appended end by end. With each input being a 98 frame spectrogram, the final 5-second input was 490 frames. It is important to note that creating a spectrogram from a 5-second audio clip will yield 498 frames. However, combining five 1-second spectrograms will give 490 frames. This is due to windowing and various other settings used by the algorithm for the spectrogram.

To enhance robustness, each 1s keyword segment underwent a multi-stage augmentation pipeline following Tang et al. \cite{tang2017}: a uniform ±100 ms temporal shift; an 80\% chance of adding MUSAN noise at a random 0–15 dB SNR; a 30\% chance of pitch-shifting by ±1–2 semitones; and convolutional reverb using room-impulse responses recorded with the target device’s microphone. After conversion to 32-bin Mel-spectrograms, SpecAugment \cite{SpecAugment} was applied with two time masks (20 frames) and two frequency masks (7 bins).

To model out-of-vocabulary and noise conditions \cite{warden2018}, two classes were added: UNKNOWN (LibriSpeech clips without target keywords) and NEGATIVE (10,000 1-second MUSAN noise/ambient clips \cite{snyder2015}). This expanded the set from 49 to 51 classes. From an initial $\sim$471K one-second clips, each class was downsampled to 20K examples to reduce imbalance, yielding $\sim$440K clips and $\sim$88K 5-second spectrograms for training and evaluation.

\subsubsection{Training Procedure}

% Three architectures were evaluated in this work:  
% \begin{itemize}
%   \item \textbf{DSResNet‐SE:} A small‐footprint ResNet built from depthwise‐separable residual blocks augmented with Squeeze‐and‐Excitation (SE) modules.  It comprises four stages of ResDS+SE layers, interleaved with pooling, and ends with a global average pool and 51‐way softmax (Table~\ref{tab:model_architecture}).  The network contains approximately 218k trainable parameters and requires $\sim$105M MACs.  
%   \item \textbf{DS‐CNN:} A depthwise‐separable CNN (MobileNet‐style) that factorizes each standard convolution into a depthwise pass followed by a pointwise projection.  This design achieves parameter efficiency (230k parameters) and $\sim$50M MACs while retaining strong accuracy on keyword‐spotting tasks.  
%   \item \textbf{TeNet:} A temporal‐efficient network built around inverted bottleneck blocks that apply 1D depthwise convolutions along the time axis, with residual shortcuts for fusion.  TeNet has 158k parameters and incurs a higher compute cost ($\sim$340M MACs) due to its long receptive fields in time.  
% \end{itemize}

The DSResNet-SE model was compared with the depthwise-separable CNN (DS-CNN) model \cite{zhang2018} and the Temporal Efficient neural network TENet by Li et al. \cite{li2020} to evaluate the proposed model's effectiveness. The DS-CNN model is made up of depthwise-separable convolutions for parameter efficiency. The TENet model is built around inverted bottleneck blocks that are made up of depthwise convolutions with a residual connection. Its main function is to perform convolutions along the time axis. To keep comparisons fair, the model parameters were kept within similar ranges. The proposed DSResNet-SE model has 218K trainable parameters, and comprises four stages of ResDS+SE layers interleaved with pooling, and ends with a global average pool and 51-way softmax (Table \ref{tab:model_architecture}).

% Three architectures were evaluated, DSResNet-SE, DSCNN, and TeNet. The DSResNet-SE model comprises four stages of ResDS+SE layers interleaved with pooling, and ends with a global average pool and 51-way softmax ; the network contains approximately 218k trainable parameters and requires $\sim$105M MACs. DSCNN is a MobileNet-style depthwise-separable CNN with 230k parameters and $\sim$50M MACs. TeNet is a temporal-efficient network built around inverted bottleneck blocks that apply 1D depthwise convolutions along the time axis, with residual shortcuts for fusion; it uses 158k parameters but incurs a higher compute cost ($\sim$340M MACs) due to its long receptive fields in time.

All models were trained in TensorFlow on an a $80\%/10\%/10\%$ stratified split of the $\sim88$K five‐second segments. $\sim71$K for training, $\sim8$K for validation, and $\sim9$K for final testing. The stratified split preserved the distribution of all 51 classes. To estimate variability, the training was repeated for three independent trials with distinct random seeds, and metrics are reported as the mean \(\pm\) standard deviation over these runs.

\begin{table}[H]
    \centering
    \scriptsize
    \begin{tabular}{lcccc}
        \hline
        Input                & Operator                & Channels     & Stride & Param.  \\ \hline
        32 x 490 x 1         & Conv2D 3x3, BN, ReLU6 & 32           & (1,1)  & 169     \\
        32 x 490 x 32        & MaxPool2D 2x2           & 32           & (2,2)  & 0       \\
        16 x 245 x 32        & ResDS + SE              & 64           & (1,1)  & 10,404  \\
        16 x 245 x 64        & MaxPool2D 2x2           & 64           & (2,2)  & 0       \\
        8 x 122 x 64         & ResDS + SE              & 128          & (1,1)  & 38,216  \\
        8 x 122 x 128        & MaxPool2D 2x2           & 128          & (2,2)  & 0       \\
        4 x 61 x 128         & ResDS + SE              & 256          & (1,1)  & 146,064 \\
        4 x 61 x 256         & MaxPool2D 2x2           & 256          & (2,2)  & 0       \\
        2 x 30 x 256         & AvgPool2D 1x6           & 256          & (1,6)  & 0       \\
        2 x 5 x 256          & Conv2D 2x1, Softmax    & 51           & (1,1)  & 26,163  \\
        1 x 5 x 51           & Reshape                 & -            & -      & 0       \\ \hline
    \end{tabular}
    \caption{DSResNet-SE Architecture.}
    \label{tab:model_architecture}
\end{table}

Adam optimiser with an initial learning rate of $2\times10^{-3}$ and batch size of 64 was used. Class imbalance was addressed by weighting each example inversely to its class frequency. An adaptive learning‐rate scheduler monitored validation loss and halved the learning rate after three stagnant epochs.  Early stopping on validation loss (patience = 5) halted training well before the 100‐epoch limit that was set. All per‐epoch loss and accuracy values were logged via \texttt{CSVLogger}. To evaluate robustness on truly unseen data, the best model from each trial was also tested (without fine‐tuning or any type of 'cheating') on the out‐of‐domain IEMOCAP corpus, providing an end‐to‐end measure of real‐world efficacy.

% To avoid over-fitting and unnecessary compute, an early stopping mechanism is employed on the validation loss with a patience of 5 epochs.  Training halts as soon as the validation loss stagnates. All per-epoch metrics (training/validation loss and accuracy) are written to a training log via a \texttt{CSVLogger} for downstream plotting and analysis. Although the limit is up to 300 epochs, early stopping typically triggers before it reaches that point, at which point the model has converged on both the training and validation sets. 

% Costly cross-validation routines were avoided in favor of a single held-out validation set (10\%) for monitoring generalization during training and a separate unseen test split (10\%) for final performance assessment. Additionally, the model was evaluated on the entirely out-of-domain IEMOCAP corpus, without any fine-tuning or data leakage. Meaning that the model's performance on IEMOCAP will serve as an accurate measure of the model’s real-world efficacy.

\subsubsection{Evaluation}

In Table \ref{tab:keyword_all_results}, DSResNet-SE consistently outperforms the other models. The Word Error Rate (WER) measures the fraction of words incorrectly recognised, it is computed as $\text{WER}=\frac{S+D+I}{N}$, where $S$=substitutions, $D$=deletions, $I$=insertions, and $N$=total words in the ground truth. Substitutions are when the prediction includes an incorrect word, deletions are when words are omitted (nothing was predicted), and insertions are extra words not present in the ground truth (extra prediction). Two new metrics are introduced for further analysis: the False Alarm Rate (FAR), which details the proportion of negative events that the model flags as positive, and the Miss Rate (MR) which denotes the proportion of positive events that the model fails to detect.

\begin{table}[H]
    \centering
    \scriptsize
    \begin{tabular}{cccc}
        \hline
        \textbf{Metrics} & \textbf{DSResNet-SE} & \textbf{DS-CNN} & \textbf{TeNet} \\ \hline
        {Accuracy}& 89.74 $\pm$ 0.17 & 67.30 $\pm$ 0.50 & 57.30 $\pm$ 0.32 \\
        % {F1 Score} & 89.72 $\pm$ 0.17& 67.19 $\pm$ 0.49& 56.87 $\pm$ 0.33\\
        {F1 Score} & 0.8972 $\pm$ 0.0017 & 0.6719 $\pm$ 0.0049 & 0.5687 $\pm$ 0.0033 \\
        {WER} & 10.26 $\pm$ 0.17 & 32.70 $\pm$ 0.50 & 42.70 $\pm$ 0.32 \\
        {FAR} &  1.17  $\pm$ 0.04 & 2.17 $\pm$ 0.11 & 3.11 $\pm$ 0.05 \\
        {MR} & 1.36  $\pm$ 0.07 & 4.63 $\pm$ 0.30 & 4.04 $\pm$ 0.03 \\ \hline
    \end{tabular}
    \caption{Model results on LibriSpeech test set. All results are in \%, except for the F1 score.}
    \label{tab:keyword_all_results}
\end{table}

% \begin{wraptable}{r}{0.6\textwidth}  % 'r' = right side, width = 0.5\textwidth
%     \centering
%     \scriptsize
%     % \setlength\tabcolsep{4pt}
%     % \renewcommand{\arraystretch}{0.85}
%     \begin{tabular}{cccc}
%         \hline
%         \textbf{Metrics} & \textbf{DSResNet-SE} & \textbf{DS-CNN} & \textbf{TeNet} \\ \hline
%         {Accuracy}& 89.74 $\pm$ 0.17 & 67.30 $\pm$ 0.50 & 57.30 $\pm$ 0.32 \\
%         % {F1 Score} & 89.72 $\pm$ 0.17& 67.19 $\pm$ 0.49& 56.87 $\pm$ 0.33\\
%         {F1 Score} & 0.8972 $\pm$ 0.0017 & 0.6719 $\pm$ 0.0049 & 0.5687 $\pm$ 0.0033 \\
%         {Word Error Rate}& 10.26 $\pm$ 0.17 & 32.70 $\pm$ 0.50 & 42.70 $\pm$ 0.32 \\
%         {False Alarm Rate}&  1.17  $\pm$ 0.04 & 2.17 $\pm$ 0.11 & 3.11 $\pm$ 0.05 \\
%         {Miss Rate}& 1.36  $\pm$ 0.07 & 4.63 $\pm$ 0.30 & 4.04 $\pm$ 0.03 \\ \hline
%     \end{tabular}
%     \caption{Model results on LibriSpeech test set. All results are in \%, except for the F1 score.}
%     \label{tab:keyword_all_results}
% \end{wraptable}

On LibriSpeech, DSResNet-SE outperforms both baselines, achieving 89.74\% accuracy with an F1 score of 0.8972, a word-error rate of 10.26\%, and keeping false-alarm and miss rates to just 1.17\% and 1.36\%, respectively. By contrast, DS-CNN—while still attractive for on-device use—drops to 67.30\% accuracy (F1 = 0.6719), with a 32.70\% WER, 2.17\% false-alarm rate, and 4.63\% miss rate. TeNet lags behind both convolutional models, at only 57.30\% accuracy (F1 = 0.5687), a 42.70\% WER, and miss/false-alarm rates above 3\%.

After training, the models are evaluated on the IEMOCAP dataset, which was not used at all in the training process. This is to assess true out-of-distribution performance and to quantify how well the KWS models generalise beyond the controlled keywords they were trained on. Unlike the clean, read-speech segments in LibriSpeech, IEMOCAP contains both scripted and fully improvised emotional dialogues.

In the scripted IEMOCAP sessions (Table \ref{tab:kws_results_iemocap}), DSResNet-SE achieved the best balance with an F1 of 0.8322 (precision 0.9307, recall 0.7553), corresponding to WER 28.13\%, FAR 6.93\%, and MR 24.47\%. DS-CNN, despite very high precision (0.9421), suffered from low recall (0.3337), yielding an F1 of 0.4899 and WER 67.28\%, while TeNet’s near-zero recall (0.0808) gave it an F1 of only 0.1466.

Across the improvised sessions, all models declined in performance but retained their relative order: DSResNet-SE remained strongest, DS-CNN’s F1 fell to 0.2764 (precision 0.7577, recall 0.1750), and TeNet performed at chance. When pooling both scripted and improvised data, DSResNet-SE kept the lead with F1 0.6476 (precision 0.8017, recall 0.5669), WER 48.35\%, FAR 18.89\%, MR 43.31\%; DS-CNN held an F1 of 0.3368 (precision 0.8099, recall 0.2199); and TeNet remained unsuitable (F1 0.0843, recall 0.0463, WER 95.39\%). 

\begin{figure}[H]
  \centering
  \begin{subfigure}[b]{0.45\textwidth}
    \includegraphics[width=0.85\textwidth]{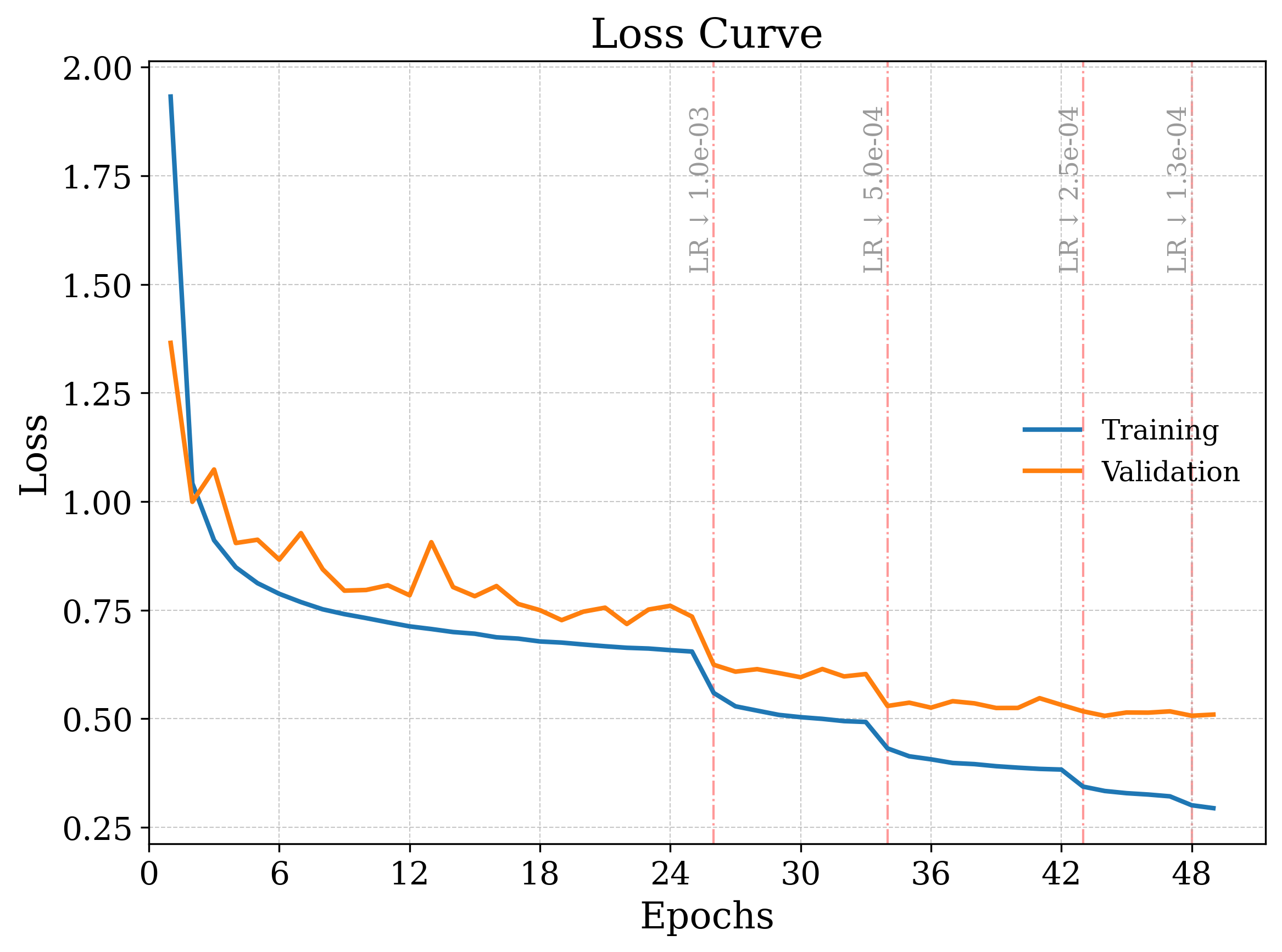}
    \caption{Training Loss Curve}
    \label{fig:plot1}
  \end{subfigure}
  \hfill
  \hfill
  \\
  \begin{subfigure}[b]{0.45\textwidth}
    \includegraphics[width=0.85\textwidth]{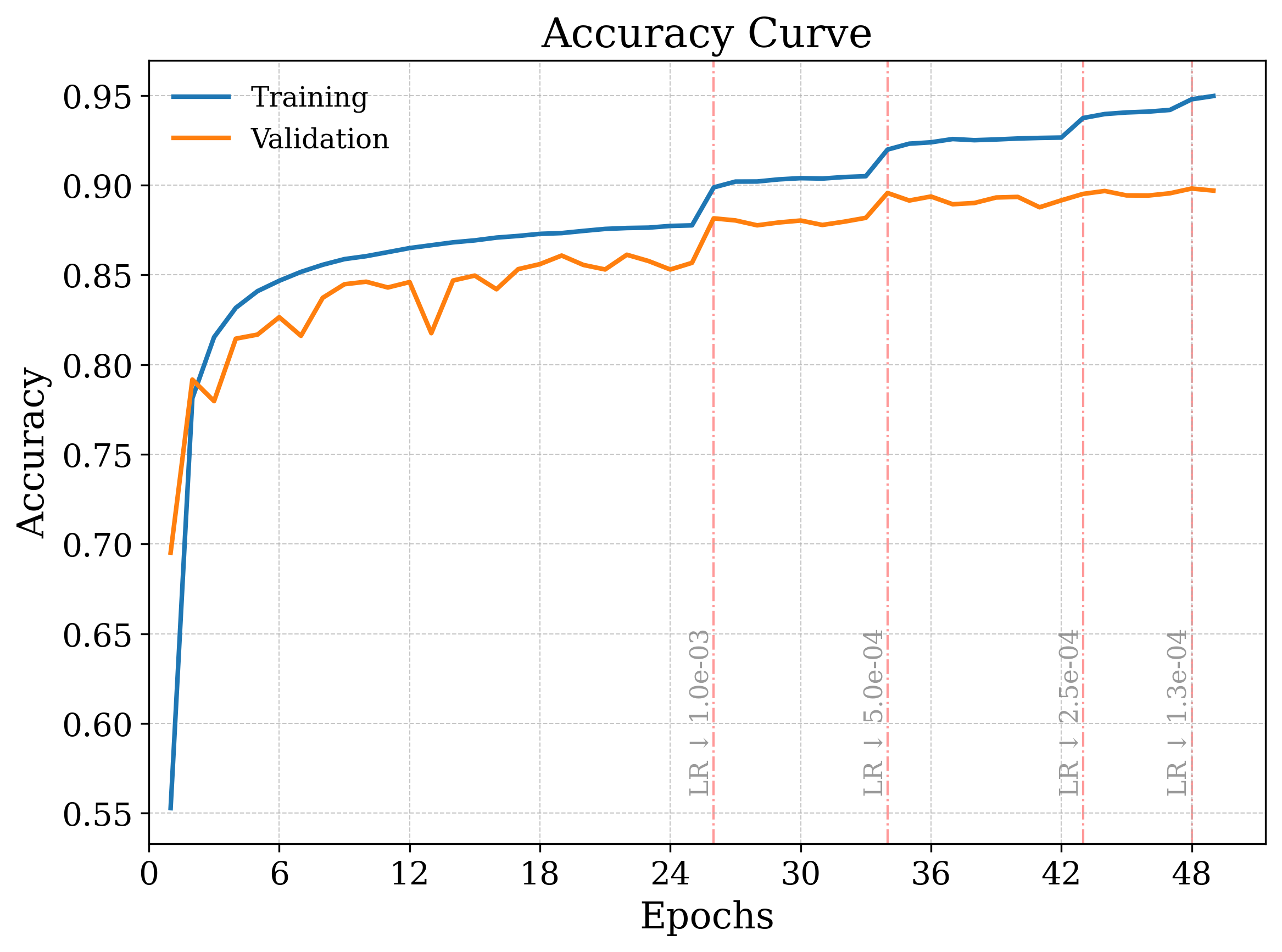}
    \caption{Training Accuracy}
    \label{fig:plot2}
  \end{subfigure}
  \caption{Training loss and accuracy of the DSResNet-SE model's best trial run. The red horizontal dashed line represents the epoch where the learning rate decayed.}
  \label{fig:model_training_results}
\end{figure}

\begin{table*}[ht]
\centering
% \footnotesize
\begin{tabular}{cccccccc}
\hline
                                                                              &             & \textbf{F1}     & \textbf{Precision} & \textbf{Recall} & \textbf{WER (\%)} & \textbf{FAR (\%)} & \textbf{MR (\%)} \\ \hline
\multirow{3}{*}{\begin{tabular}[c]{@{}c@{}}IEMOCAP\\ Scripted\end{tabular}}   & DSResNet-SE & \textbf{0.8322} & 0.9307             & \textbf{0.7553} & \textbf{28.13}    & 6.929             & \textbf{24.47}   \\
                                                                              & DSCNN       & 0.4899          & \textbf{0.9421}    & 0.3337          & 67.28             & \textbf{5.788}    & 66.63            \\
                                                                              & TeNet       & 0.1466          & 0.9091             & 0.0808          & 91.99             & 9.087             & 91.92            \\ \hline
\multirow{3}{*}{\begin{tabular}[c]{@{}c@{}}IEMOCAP\\ Improvised\end{tabular}} & DSResNet-SE & \textbf{0.5747} & 0.7507             & \textbf{0.4925} & \textbf{56.34}    & 23.61             & \textbf{50.75}   \\
                                                                              & DSCNN       & 0.2764          & \textbf{0.7577}    & 0.1750          & 82.74             & \textbf{18.96}    & 82.50            \\
                                                                              & TeNet       & 0.0598          & 0.4206             & 0.0327          & 96.73             & 21.10             & 96.73            \\ \hline
\multirow{3}{*}{Combined}                                                     & DSResNet-SE & \textbf{0.6476} & 0.8017             & \textbf{0.5669} & \textbf{48.35}    & 18.89            & \textbf{43.31}    \\
                                                                              & DSCNN       & 0.3368          & \textbf{0.8099}    & 0.2199          & 78.37             & \textbf{15.23}    & 78.00            \\
                                                                              & TeNet       & 0.0843          & 0.5588             & 0.0463          & 95.39             & 17.70             & 95.37            \\ \hline
\end{tabular}
\caption{Model evaluation on the IEMOCAP's scripted and improvised recording sessions. FAR is the False Alarm Rate, and MR is the Miss Rate. The best performing model from the three trials was picked for evaluation.}
\label{tab:kws_results_iemocap}
\end{table*}

Comparing DSResNet-SE's performance to the LibriSpeech results, it can be seen that the F1 score remained similar, decreasing from 0.8972 to 0.8322. This indicates that the enhancements and augmentations made to the training data were sufficient to help the model generalise to unseen data, with a small residual domain mismatch. LibriSpeech’s clean audiobook recordings lack the spontaneous, conversational dynamics present in the improvised sessions, which explains the drop in performance on the improvised dataset.

% On the improvised recordings, the performance gap widened.  DSResNet‐SE’s F1 fell to \(0.5747\) (precision \(0.7507\), recall \(0.4925\)), with WER \(=56.34\%\), FAR \(=23.61\%\) and MR \(=50.75\%\).  DS‐CNN managed an F1 of \(0.2764\) (precision \(0.7577\), recall \(0.1750\)) but incurred a WER of \(82.74\%\) and MR \(=82.50\%\).  TeNet again failed to generalize (F1 \(=0.0598\), recall \(0.0327\), WER \(=96.73\%\)).

% When scripted and improvised sessions were pooled, DSResNet‐SE delivered the best overall balance: F1 \(=0.7895\), precision \(0.7068\), recall \(0.9152\), WER \(=45.15\%\), FAR \(=29.31\%\) and MR \(=8.48\%\).  DS‐CNN trailed with F1 \(=0.3368\), precision \(0.8099\), recall \(0.2199\), WER \(=78.37\%\) and MR \(=78.00\%\).  TeNet performed worst (F1 \(=0.0843\), recall \(0.0463\), WER \(=95.39\%\)).

% \begin{figure}[H]
%     \centering
%     \includesvg[width=0.8\linewidth]{keyword/figures/F1_plot.svg}
%     \caption{F1 Plot}
%     \label{fig:F1_results}
% \end{figure}

\begin{table}[H]  % 'r' = right side, width = 0.5\textwidth
  \centering
  \scriptsize
  \begin{tabular}{l r r}
  \hline
    \textbf{Model}      & \textbf{Parameters} & \textbf{MACs} \\ \hline
    DSResNet-SE & 218\,K & 105\,M \\
    DS-CNN      & 230\,K &  50\,M \\
    TeNet       & 158\,K & 340\,M \\ \hline
  \end{tabular}
  \caption{Model complexity. MACs were obtained using the \texttt{keras\_flops} library}
  \label{tab:complexity_accuracy}
\end{table}

% \begin{wraptable}{r}{0.4\textwidth}  % 'r' = right side, width = 0.5\textwidth
%   \centering
%   \scriptsize
%   \begin{tabular}{l r r}
%   \hline
%     \textbf{Model}      & \textbf{Parameters} & \textbf{MACs} \\ \hline
%     DSResNet-SE & 218\,K & 105\,M \\
%     DS-CNN      & 230\,K &  50\,M \\
%     TeNet       & 158\,K & 340\,M \\ \hline
%   \end{tabular}
%   \caption{Model complexity. MACs were obtained using the \texttt{keras\_flops} library}
%   \label{tab:complexity_accuracy}
% \end{wraptable}

As shown in Table \ref{tab:complexity_accuracy}, TeNet, despite its lower parameter count, scales very poorly. Its multiply–accumulate operations (MACs) requirements grow steeply with input length, resulting in excessive latency and energy draw. In contrast, DS-CNN exhibits the best scaling behaviour, incurring the fewest additional MACs per extra parameter and thus delivering the fastest inference and lowest power consumption. However, its baseline detection performance remains relatively low. DSResNet-SE occupies the middle ground. The addition of SE blocks introduces a moderate compute overhead but yields substantial accuracy gains, enabling balanced scaling of throughput, power use, and robustness. Overall, DSResNet-SE offers the most practical trade-off for on-device keyword spotting, with DS-CNN as a viable low-power fallback.

\subsection{Emotion Classification Model} % Here shashwat!
\subsubsection{Dataset Preprocessing}
To develop a real-time speech emotion recognition (SER) model suitable for deployment on the Coral Dev Board Micro, a comprehensive preprocessing of the IEMOCAP dataset \cite{iemocap} was done. The preprocessing strategy was designed to optimize computational efficiency ensuring optimal performance on resource-constrained MCU hardware.

Each audio file from the IEMOCAP dataset was segmented into 5-second chunks with a 1-second overlap. This approach provides a balance between temporal resolution and computational load, facilitating real-time inference on embedded systems. Overlapping windows help capture transitional emotional cues and increase the dataset's size, which is beneficial for training deep learning models \cite{trigeorgis2018speech}.

The IEMOCAP recordings feature two channels per session, corresponding to male and female speakers. To enhance the model's ability to generalize across different speaker genders, each audio chunk was duplicated, isolating the male channel in one copy and the female channel in the other. Although the non-target speaker's voice remains faintly audible, this approach increases data diversity and encourages the model to focus on the dominant speaker's emotional cues. This method aligns with findings that gender-specific features can improve SER performance \cite{zhang2023deep}.

For classification,  only five emotion categories were retained: \textit{neutral}, \textit{happy}, \textit{sad}, \textit{angry}, and \textit{excited}. All other emotion classes were discarded. Additionally, the \textit{happy} and \textit{excited} classes were merged into a single class, as is common practice in the SER literature \cite{Cai2021, Wu2019}. This simplification not only helps to address class imbalance and improve model robustness but also aligns our evaluation with other state-of-the-art benchmarks that use the same five-class setting on the IEMOCAP dataset \cite{Padi2022}. Also, rather than using categorical labels, soft-labels were created by determining the fraction of the annotations each emotion received in the clip. This choice addressed the ambiguity in labelling when multiple emotions were present, or when the assessment of independent annotators was inconsistent. 

To enable the model to recognize the absence of speech or neutral background conditions, a 'none' class was introduced, similar to the KWS model's 'UNKNOWN' class. This class comprises of background noise samples taken from the MUSAN dataset, representing various ambient environments. Segments from the IEMOCAP dataset where only the non-target speaker is active were also taken, resulting in low-amplitude speech. This strategy trains the model to distinguish between active speech and silence or background noise, enhancing its real-world applicability. Incorporating non-speech segments is a common practice to improve the robustness of SER systems in practical scenarios \cite{schuller2022evaluating}.

For model evaluation, Session 1 of the IEMOCAP dataset was designated as the validation set, while Sessions 2 through 5 served as the training set. This session-based split ensures speaker independence between training and validation data, providing a robust assessment of the model's generalization capabilities. To assess the model's performance in real-world conditions, the entirety of Session 1 was recorded using the Coral Dev Board Micro's onboard microphone. As these recordings were mono (one channel), the audio was segmented into 5-second chunks and the segments containing overlapping speech from both speakers were excluded. This real-world validation set guided hyperparameter tuning and final model selection, ensuring alignment with deployment scenarios. Moreover, background-noise clips recorded directly on the microcontroller were distributed evenly across the sessions. Evaluating models on data captured from the target deployment environment is crucial for assessing real-time performance and robustness \cite{rehman2022real}.

\subsubsection{Training Procedure}
Models were trained for $125$ epochs using the Adam optimiser with initial learning rate $1 \times 10^{-4}$ and batch size $32$. These hyperparameters were optimised independently by iterating over a selection of possible values. For each training run, the model configuration with the highest macro-averaged F1-score on the microcontroller validation set was selected. Several methods were applied to reduce overfitting, specifically $L2$-regularisation with weight decay $1 \times 10^{-5}$, dropout at a rate of $0.1$ and the use of the cosine decay learning rate scheduler, which reduced the learning rate by a factor of $2$ over the course of training. 

A weighted cross-entropy loss was used, which assigned additional importance to under-represented categories. Additionally, samples recorded on the microcontroller were weighted by a factor of $1.1$, encouraging models to perform better on the microcontroller.

Prior to spectrogram creation, several augmentations were applied to the waveforms, similarly to those described in section \ref{kw_preprocessing}. The methods applied were as follows: Gaussian noise addition with $\sigma=5 \times 10^{-3}$ \cite{augmentation_overview}, the inclusion of randomly sampled background noise from the \texttt{MUSAN} dataset \cite{snyder2015} and convolving waveforms with room impulse response clips \cite{rir}. Upon spectrogram creation, SpecAugment \cite{SpecAugment} was applied, using frequency masks up to 4 bins wide and time masks of at most 50 frames. The augmentations occurred with probabilities 0.15, 0.2, 0.1, 0.2 and 0.2 respectively - values which were obtained by sequential tuning; gradually increasing each parameter until no improvement in performance was observed.

\subsubsection{Ablation Study}

To evaluate the contribution of each component to the model's performance, an ablation study was conducted. The primary goal was to understand the effect of augmentations, recorded MCU validation samples in training, and fusion with the keyword spotting model (KWS).
Each experiment builds on the previous and introduces one change at a time to isolate its effect. 

In Experiment 1, the model was trained without any of our proposed waveform or spectrogram augmentations and without integrating the keyword-spotting (KWS) branch; this established a reference point for baseline performance on MCU-recorded validation data. Building on that, Experiment 2 incorporated all augmentation strategies (Gaussian noise, MUSAN background, RIR convolution, and SpecAugment) while still omitting the KWS module, demonstrating how data augmentation alone boosts generalisation. To test whether the micro validation set (that was recorded on the MCU) would give a performance boost or not, they were added in training. In Experiment 3, a subset of the recorded microcontroller validation clips was removed from validation and kept away to evaluate the model performance on the remaining validation set. Then in Experiment 4, that same subset was added into the training to quantify the gains from including representative MCU audio during training. Next, Experiment 5 added the keyword-spotting network via an early-fusion architecture. KWS model's output embeddings were given to the acoustic branch as an input to the encoder to measure its contribution. Finally, Experiment 6 employed late fusion, by concatenating KWS embeddings with the embeddings of the encoder before the classifier head; this configuration achieved the highest Micro Validation F1, indicating that decoupling the two tasks and merging their output embeddings produces the most robust performance on the MCU validation set. The results in terms of best achieved Micro F1-score for each configuration are reported in Table~\ref{tab:ablation}.

% \begin{table}[h!]
% \centering
% \begin{tabular}{|c|l|c|}
% \hline
% \textbf{Exp. No.} & \textbf{Configuration}                   & \textbf{Best Micro F1} \\ 
% \hline
% 1                 & No augmentation, no KWS                 & 0.3366                 \\ 
% 2                 & With augmentation, no KWS               & 0.4986                 \\ 
% 3                 & Micro val subset removed                & 0.3999                 \\ 
% 4                 & Micro val subset added in train         & 0.5477                 \\ 
% 5                 & Early-fusion KWS model                  & 0.5449                 \\ 
% 6                 & Late-fusion KWS model                   & \textbf{0.6107}        \\ 
% \hline
% \end{tabular}
% \caption{Ablation study results on recorded microcontroller validation set.}
% \label{tab:ablation}
% \end{table}
\begin{table}[ht]
    \centering
    \footnotesize
    \begin{tabular}{clc}
        \toprule
        \textbf{Exp. No.} & \textbf{Configuration} & \textbf{Best Micro F1} \\
        \midrule
        1 & No augmentation, no KWS                 & 0.3366 \\
        2 & With augmentation, no KWS               & 0.4986 \\
        3 & Micro Val subset removed                & 0.3999 \\
        4 & Micro Val subset added in train         & 0.5477 \\
        5 & Early-fusion with KWS model                  & 0.5449 \\
        6 & Late-fusion with KWS model                   & \textbf{0.6107} \\
        \bottomrule
    \end{tabular}
    \caption{Ablation study results on recorded microcontroller validation set.}
    \label{tab:ablation}
\end{table}

The results demonstrate the cumulative importance of each enhancement. Augmentations provided a substantial boost over the baseline. Including recorded microcontroller data in training helped significantly. While early fusion of the KWS model improved performance, late fusion outperformed all setups.

\subsubsection{Evaluation}

To comprehensively assess both classification effectiveness and edge deployment viability, the following metrics were evaluated for our best-performing configuration (Experiment 6: Late-fusion Model). This setup achieved a best micro validation F1 of 0.6107 with an accuracy of 0.6438, and a best validation F1 of 0.5256 with an accuracy of 0.4913.

% \begin{table}[h!]
% \centering
% \begin{tabular}{|l|p{8cm}|}
% \hline
% \textbf{Metric}            & \textbf{Description / Value}                                                                                  \\ \hline
% Macro F1                  & 0.6107 \\ \hline
% Weighted F1               & 0.644 \\ \hline
% Accuracy                  & 0.6438 \\ \hline
% Memory Footprint          & Peak RAM usage during inference—ensures suitability for microcontroller constraints. \\ \hline
% Model Size                & Flash storage size of the quantized model binary—reflects on-device storage requirements. \\ \hline
% Energy per Inference      & Estimated energy consumed per inference—critical for battery- or energy-constrained edge devices. \\ \hline
% \end{tabular}
% \caption{Evaluation metrics of the Final Model}
% \label{tab:eval_metrics}
% \end{table}
\begin{table}[ht]
    \centering
    \small
    \begin{tabular}{lc}
        \toprule
        \textbf{Metric} & \textbf{Value} \\
        \midrule
        Macro F1             & 0.6107 \\
        Weighted F1          & 0.6440 \\
        Accuracy             & 0.6438 \\
        Num. of Parameters   & 734,261 \\
        MCU model size       & 1.8 MB \\
        Memory Footprint     & 1.4 MB \\
        \bottomrule
    \end{tabular}
    \caption{Evaluation metrics of the Final Model}
    \label{tab:eval_metrics}
\end{table}

\begin{figure}[H]
    \centering
    \includegraphics[width=\linewidth]{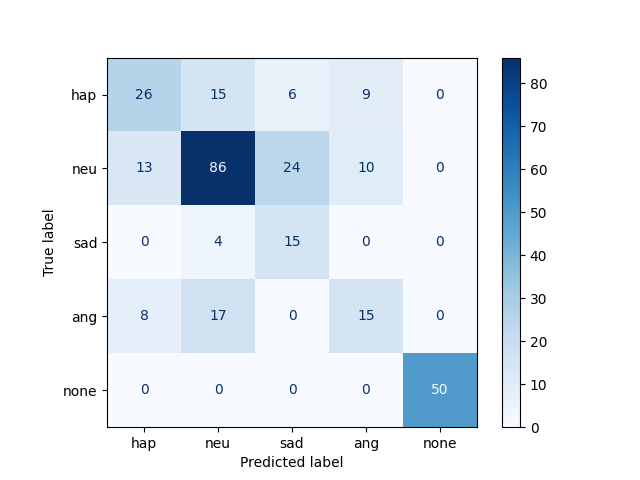}
    \captionof{figure}{Confusion Matrix on the validation set}
    \label{fig:val_results}
\end{figure}

Table~\ref{tab:class_metrics} presents the per-class precision, recall, and F1-score on the microcontroller validation set. Alongside it, Figure~\ref{fig:val_results} shows the confusion matrix of the best performing model.

\begin{table}[h!]
    \centering
    \begin{tabular}{lccc}
        \toprule
        \textbf{Class} & \textbf{Precision} & \textbf{Recall} & \textbf{F1-score} \\
        \midrule
        Happy      & 0.5532 & 0.4643 & 0.5049 \\
        Neutral    & 0.7049 & 0.6466 & 0.6745 \\
        Sad        & 0.3333 & 0.7895 & 0.4688 \\
        Angry      & 0.4412 & 0.3750 & 0.4054 \\
        No Emotion & 1.0000 & 1.0000 & 1.0000 \\
        \midrule
        \textbf{Macro avg} & 0.6065 & 0.6551 & 0.6107 \\
        \bottomrule
    \end{tabular}
    \caption{Per-class and macro-averaged metrics on the microcontroller validation set}
    \label{tab:class_metrics}
\end{table}

% \end{minipage}

% insert training here
% \begin{itemize}
%     \item Hyperparameters, training epochs, and loss functions
%     \item Evaluation metrics for keyword spotting (accuracy, precision, recall) and sentiment analysis
% \end{itemize}
% \subsubsection{Evaluation}
% \begin{itemize}
%     \item Performance metrics for the combined model (accuracy, F1-score on sentiment classification)
%     \item Discussion on the benefit of multi-modal fusion
% \end{itemize}

% \subsection{Baseline Comparisons and Ablation Studies}
% \begin{itemize}
%     \item Analysis of the effects of quantization and model freezing on performance
% \end{itemize}

%%%%%%%%%%%%%%%%%%%%%%%%%%%
%%%%%%%%%%%%%%%%%%%%%%%%%%%

\section{Edge Deployment and Quantization}

In the next stage of the workflow, the Coral Micro’s Edge TPU is utilized. This purpose-built, low-power device delivers up to 4 TOPS (trillions of operations per second) while consuming under 0.5 W per TOPS \cite{coral_performance_benchmarks}. Only 8-bit integer operations are supported on the Edge TPU, requiring all weights and activations to be quantized in advance (done via TensorFlow Lite’s post-training quantization). A limited list of operations are supported, which has driven architectural adaptations such as the adoption of ReLU6 activations and depthwise-separable layers. Any unsupported layers or data types must be rewritten or approximated using TPU-compatible equivalents.

\subsection{Impact on models and data pre-quantization} \label{edgetpu_limitations}
% \texttt{ReLU6} activation functions were selected throughout the model in order to maximise post-quantization performance \cite{Krizhevsky2010}.

% \subsection{Architecture Adaptations} \label{edgetpu_limitations}
% Due to the restrictions imposed by the Edge TPU on the available TensorFlow operations \cite{coral2025edgetpu}, several adaptations were made to the model architecture to enable effective deployment on the Coral Microcontroller.

The Edge TPU is limited to processing unbatched inputs and supports only one-, two-, or three-dimensional tensors. When a tensor exceeds three dimensions, only its three innermost dimensions may have sizes greater than one \cite{coral2025edgetpu}. These restrictions led the KWS model to consume the entire 5-second spectrogram in a single pass, emitting one label per second, rather than batching five 1-second spectrograms. In the emotion classification model, to constrain the \(O(N^2)\) complexity of attention on a \(32\times498\) Mel-spectrogram (transposed into a sequence of 498 tokens), a CNN reduces the temporal dimension from 498 to 32 frames, to preserve local temporal features, before the transformer. 

As the TPU’s on‐chip SRAM can cache only about 8 MB of model parameters (after reserving space for the inference engine), compiled models must stay below this limit to be fully resident and accelerated on the Edge TPU, which was achieved by the model as seen in Table \ref{tab:eval_metrics}. The dimensionality of the feed-forward blocks in the emotion classification model was kept constant rather than expanding by a factor of four, as in standard transformer architectures \cite{transformer}, to meet the Edge TPU’s memory and compute budgets. Not to mention, the spectral resolution of the Mel-spectrograms were halved, decreasing the channel count was from 64 to 32, to conform to the TPU’s 8 MB on-chip memory limit.

\subsection{Deployment Overview and Considerations}
% Deployment of TensorFlow trained models on embedded systems requires conversion to TensorFlow-Lite (\texttt{TFLite}) format, which quantises models to INT8 precision \cite{coral2025edgetpu} as seen in Figure~\ref{fig:flow-combined}.

% In TFlite, model architectures are represented as computations in directed acyclic graphs composed of predefined operations such as addition (\texttt{ADD}), multiplication (\texttt{MUL}), and two-dimensional convolution (\texttt{CONV\_2D}), optimised for execution on CPUs. To enable hardware acceleration on TPUs, TFLite models must be compiled into Edge TPU-compatible formats (\texttt{tflite-for-edge-tpu}), where eligible operation sequences are merged into a single \texttt{edgetpu-custom-op}. Operations not supported by the TPU remain outside this block and when encountered, all subsequent operations are executed on the CPU, which partitions the model graph  (Figure~\ref{fig:flow1}). The compilation of models for Edge TPU imposes strict requirements; Tensor parameters must be quantized to INT8 or UINT8, tensor shapes must be statically defined, model parameters must remain constant, and tensors are restricted to 3D.

\begin{figure*}[ht]
    \centering
  \begin{subfigure}[b]{0.47\textwidth}
        \centering
        \includegraphics[width=\linewidth, height=4.5cm]{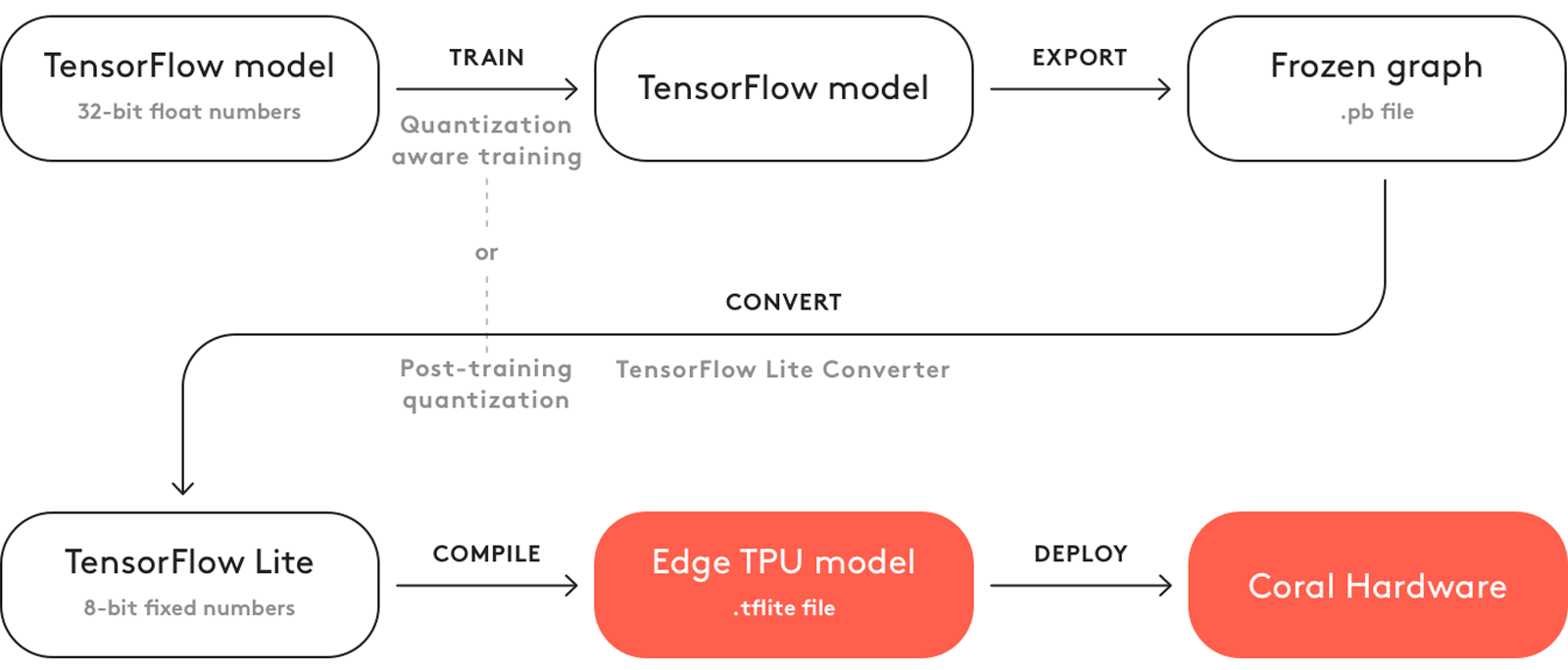}
        \caption{Workflow for converting a TensorFlow model to a TFLite-EdgeTPU binary.}
        \label{fig:flow1}
    \end{subfigure}
    \hspace{0.03\linewidth}
  \begin{subfigure}[b]{0.47\textwidth}
        \centering
        \includegraphics[width=\linewidth, height=4.5cm]{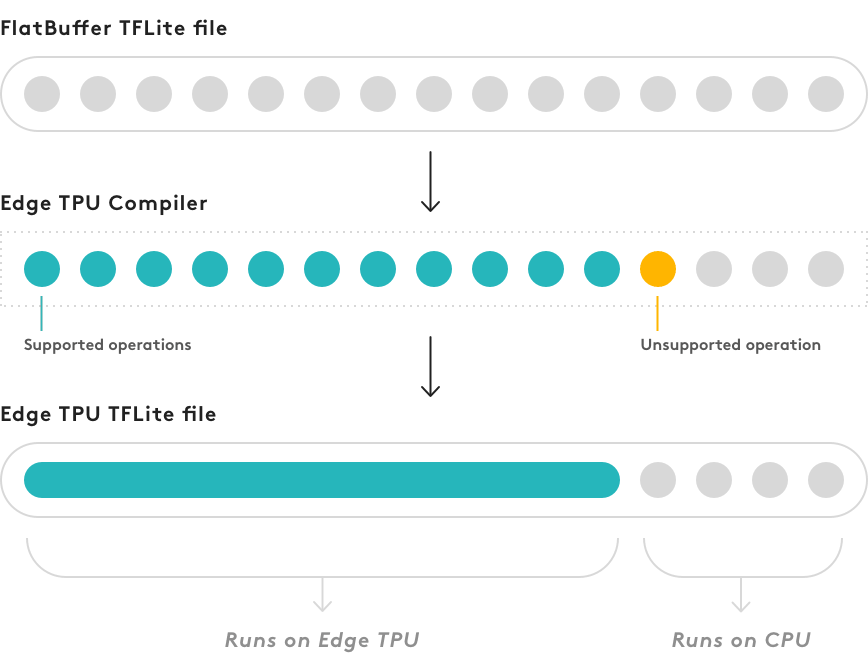}
        \caption{Edge TPU model compiler partitioning scheme \cite{coral2025edgetpu}.}
        \label{fig:flow2}
    \end{subfigure}
    \caption{Pipeline for the quantization and deployment of a TensorFlow model to the Edge TPU.}
    \label{fig:flow-combined}
\end{figure*}

The TensorFlow→TFLite→Edge TPU deployment flow is shown in Figure~\ref{fig:flow-combined}.  TensorFlow models are first converted to TFLite format and quantized to INT8 precision \cite{coral2025edgetpu}. The TFLite graph is partitioned during compiler time: supported subgraphs are merged into an \texttt{edgetpu-custom-op}, while unsupported operations remain on the CPU (Fig.~\ref{fig:flow2}). Because TPU fallbacks still execute on-device under TFLM—which supports a subset of TFLite operations—any unsupported operation both degrades performance and risks runtime failure.  Strict requirements apply: all tensor shapes must be static, parameters constant, and tensors limited to three dimensions. Compiler/runtime version compatibility is also critical for correct on-device execution \cite{coral2025edgetpu}.

On the Coral Micro, an additional constraint applies: operations that fall back to CPU execution must also be compatible with TFLM, which supports a narrower set of operations than TFLite. As a result, unsupported operations degrade inference performance and may cause runtime incompatibilities. Thus, it is critical that all operations are executed entirely on the Edge TPU. Lastly, it is important to distinguish between the TPU compiler and runtime. The compiler is an offline tool that converts TFLite models into Edge-TPU models, while the runtime operates on-device, ensuring delegated operations are executed. Runtime and Compiler version compatibility is essential for correct on-device execution\cite{coral2025edgetpu}.

\begin{figure*}[!b]
    \centering
    \includegraphics[width=1.0\linewidth]{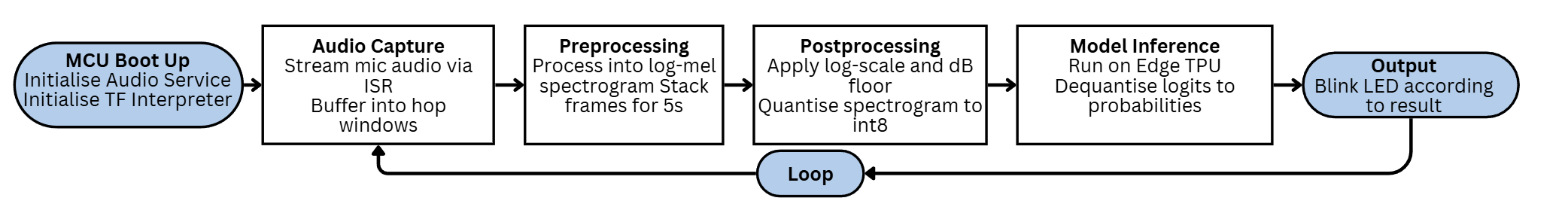}
    \caption{Flowchart illustrating the algorithm deployed on the MCU.}
    \label{fig:code-flowchart}
\end{figure*}

\subsection{Edge TPU Deployment Challenges}

% \subsubsection{Missing Operations}
% Initial model compilation failed due to absent TFLM definitions for ops like \texttt{SquaredDifference}.  Updating the runtime was not feasible due to extensive library-level incompatibilities, so missing primitives were manually reimplemented at the graph level using supported ops (e.g.\ \texttt{SUB} and \texttt{SQUARE}).
Initial model compilation failed because, unlike the compiler, the TFLM runtime lacked certain operation definitions such as \texttt{SquaredDifference}. Updating the runtime to support missing operations was deemed infeasible due to extensive library-level incompatibilities. Fixing a small issue in the massive Coral Micro repository triggers a domino effect of errors across thousands of interdependent files. Debugging it would require deep hardware and low-level systems expertise well outside the scope of the team members’ skillsets and knowledge. Ultimately, missing operators were either manually reimplemented using supported primitives, such as \texttt{SUB} and \texttt{SQUARED}, or abandoned entirely.

% % \subsubsection{Static Model Outputs}
Inference on the target model yielded identical outputs for all inputs. Deploying YAMNet confirmed the audio pipeline was correct, so the issue was traced to model shapes: fully-connected layers and multi-head attention were fed 2D/3D tensors unsupported by the Edge TPU. To bypass the size restrictions, a $1 \times 1$ convolution was implemented to conform to the TPU’s 1D–3D tensor limitation \cite{coral2025edgetpu}. A single head attention block was maintained as multi-head attention over 3D tensors was not supported. Model capacity was recovered by enlarging the CNN front end and stacking additional transformer layers, taking advantage of the memory savings from using a single head.

% Inference tests revealed invariant model outputs regardless of input audio. Deploying a known classification model (YAMNet) yielded varying probabilities, verifying correct operation of the audio pipeline and spectrogram generation. Thus, attention shifted to model architecture, where review of Edge TPU documentation indicated that fully connected layers must receive 1D inputs. Early models employed a point-wise feed forward layer (FFL) with 2D inputs, leading to invalid TPU behaviour. Similar shape issues were identified with multi-head attention blocks. For testing, the FFL was removed and a single attention head employed, which resolved the static output problem confirming full operational integrity across the entire audio-to-prediction pipeline. 

\subsection{Final Code Methodology}
Figure~\ref{fig:code-flowchart} illustrates a flowchart of the final pipeline deployed on the microcontroller. The system begins by initialising the audio frontend for preprocessing and configuring the Edge TPU interpreter with a quantised \texttt{TFLite} model. Audio data is captured in real-time via interrupt-driven streaming and buffered into overlapping hop windows. These samples are transformed into log-mel spectrogram frames and accumulated over a 5-second window. After applying log scaling and a dB floor, the spectrogram is quantized to INT8 format and passed to the model for inference. The Edge TPU processes the input and produces emotion classification logits, which are dequantised into probabilities. The predicted emotion class is then used to trigger a visual indication via LED blinking. The system resets and continues processing subsequent audio windows in real time, enabling continuous, on-device emotion recognition.

When deployed, measurements on the Coral Micro indicated an inference latency of 21–23 ms, a maximum TPU temperature of 32.5 °C, and a constant power draw of 2.5 W (5 V at 0.5 A) while executing the quantized model. On a typical 500 mAh battery pack, the microcontroller can constantly perform inference for approximately 44 minutes. In practice, continuous inference would be unsustainable. Ideally, a lightweight sound‐detection front end would trigger the full model only when relevant audio is present, substantially extending battery life. This can be achieved by utilising the Coral Micro's Dual-Cortex processor. The slower processor can be used to perform sound detection, and the faster processor reserved for the main system.

% \subsection{Quantization Techniques}
% \begin{itemize}
%     \item Description of the quantization process (post-training quantization, quantization-aware training)
%     \item Impact of quantization on model size and inference speed
% \end{itemize}

% \subsection{Impact of Quantization and Edge Deployment Metrics}
% \begin{itemize}
%     \item Comparisons in terms of latency, model size, and power consumption pre- and post-quantization
%     \item Discussion of deployment challenges and future optimizations
% \end{itemize}

% \subsection{Deployment Strategy on Edge Devices}
% \begin{itemize}
%     \item Hardware specifications and constraints
%     \item Conversion to TensorFlow Lite: challenges and optimizations
%     \item Real-world performance: latency, accuracy, and energy consumption measurements
% \end{itemize}

% \input{sentiment-analysis-model/edge tpu adaptations}

%%%%%%%%%%%%%%%%%%%%%%%%%%%
%%%%%%%%%%%%%%%%%%%%%%%%%%%

\section{Conclusion and Future Work}
% \begin{itemize}
%     \item Summary of contributions and major findings
%     \item Discussion of potential improvements (model architecture enhancements, dataset expansion, improved fusion strategies)
%     \item Future research directions (e.g., real-time adaptation, further miniaturization for edge devices)
% \end{itemize}

A modular and efficient audio–text analysis system was developed to enable real-time emotion recognition on low-power edge devices. A lightweight KWS architecture (DSResNet-SE) was proposed, incorporating residual connections, depthwise-separable convolutions, and SE blocks to balance model performance and computational efficiency. The KWS model was evaluated on both in-distribution (LibriSpeech) and out-of-distribution (IEMOCAP) datasets, where it demonstrated competitive accuracy and robustness compared to existing compact baselines. A late-fusion strategy was employed to integrate acoustic and linguistic features, further enhancing emotion-recognition performance. The final model was successfully quantized and deployed on the Coral Dev Board Micro, enabling continuous inference while maintaining user privacy.

Despite the system’s strong generalisation performance on re-recorded samples, robustness to varied environmental conditions and speaker diversity remains an area for improvement. Broader datasets and cross-domain evaluation could help address this. Future work will focus on three main areas. First, on-device personalisation techniques such as continual learning will be explored to improve adaptation to individual users. Second, integration of a lightweight sound-detection trigger will be implemented to activate inference only when relevant audio is detected, thereby extending battery life. Finally, a wearable prototype will be developed to validate portability and functionality in real-world scenarios. Through these enhancements, the framework is expected to provide a foundation for scalable, ethical, and privacy-preserving audio intelligence at the edge.

% Despite the system’s strong generalization performance on re-recorded samples, robustness to varied environmental conditions and speaker diversity remains an area for improvement. Broader datasets and cross-domain evaluation could help address this. Future work will focus on three main areas. First, on-device personalization techniques such as continual learning will be explored to improve adaptation to individual users. Second, post-quantization optimization through operator fusion and kernel tuning will be investigated to further reduce inference latency. Third, a wearable device will be prototyped to test its portability and functionality in the real-world. Through these enhancements, the framework is expected to provide a foundation for scalable, ethical, and privacy-preserving audio intelligence at the edge.

%%%%%%%%%%%%%%%%%%%%%%%%%%%
%%%%%%%%%%%%%%%%%%%%%%%%%%%

%%%%%%%%%%%%%%%%%%%%%%%%%%%
%%%%%%%%%%%%%%%%%%%%%%%%%%%

%%%%%%%%%%%%%%%%%%%%%%%%%%%
% \input{microcontroller/microntroller}
%%%%%%%%%%%%%%%%%%%%%%%%%%%

% \input{Validations/validation}

% \section{Another section}

% Final report should be 12--15 pages, \textbf{not} including the title page,
% the optional table of contents page, and the bibliography.

\clearpage
% \bibliographystyle{ieeetr}
% \bibliography{references}
% \printbibliography
\bibliographystyle{IEEEtran}
\bibliography{references} % or whatever your .bib file is called

%%%%%%%%%%%%%%%%%%%%%%%%%%%
%%%%%%%%%%%%%%%%%%%%%%%%%%%
%%%%%%%%%%%%%%%%%%%%%%%%%%%

% \appendix
% \input{appendix}

\end{document}